# A Minimal Model of the Interaction of Social and Individual learning

Kingsley J.A. Cox and Paul R. Adams

Dept Neurobiology, Stony Brook University, Stony Brook, NY 11794

Correspondence to: K.J.A. Cox   Correspondence and requests for materials should be addressed to K.J.A. Cox (email: kcox@syndar.org)

**Abstract**
Social, supervised, learning from others might amplify individual, possibly unsupervised, learning by individuals, and might underlie the development and evolution of culture. We studied a minimal model of the interaction of individual unsupervised and social supervised learning by interacting agents. Agents attempted to learn to track a hidden fluctuating "source", which, linearly mixed with other masking fluctuations, generated observable input vectors. Learning was driven either solely by direct observation of inputs (unsupervised, Hebbian) or, in addition, by observation of another agent's output (supervised, Delta rule). To enhance biological realism, the learning rules were made slightly connection-inspecific, so that incorrect learning sometimes occurs. We found that social interaction can foster both correct and incorrect learning. Useful social learning therefore presumably involves additional factors some of which we outline.

multi-agent learning; Independent Components Analysis; Social learning

**INTRODUCTION**

An important aspect of intelligence is the ability to detect regularities in complex, seemingly unpatterned, data. Brains can achieve this via learning, largely by activity-dependent adjustment of the relative strengths of synaptic connections between pairs of neurons. Since postsynaptic activity is typically caused by presynaptic activity, a learning rule operating at individual synapses that depends on both pre- and postsynaptic activity, in a broadly Hebbian manner, can detect and represent statistical regularities over an entire set of inputs, since it's driven by input correlations, both second and higher order. Learning can sometimes also be facilitated by supervisory signals, especially when a critic or teacher already has access to hidden structure.

"There are 2 contrasting but not incompatible ideas about individual human intelligence. Firstly, humans might have special brain circuits that directly underpin components of intelligence (Fodor, 1983; Pinker, 2010; Preuss et al., 1999). Second, humans may have enhanced ability to learn, either individually or from others. In particular, using symbolic communication, especially language, humans may learn about the world using guidance from other humans (Tomasello 2009), in addition to direct observation. Part of the distinctive power of human cognition, compared to other animals, might derive from the way that individual and collective learning are coupled by symbolic communication so that individual insight can spread socially, and even evolve (Richerson and Boyd, 2008) (Boyd et al., 2011) (Galef and Laland, 2005). Here we develop a minimal quantitative model of the leveraging between individual learning and social learning. Previous models of multiagent learning have been largely based on supervised learning (Sen and Weiss, 1999) (Denaro and Parisi, 1997). However, our model postulates unsupervised individual learning combined with supervised social learning from others.

Our model is based on the mathematically-transparent Independent Components Analysis (ICA) paradigm (Hyvärinen et al., 2004). In this model, input regularities are generated in the simplest possible way, by linear mixing of unstructured "causes" or "sources" (independently fluctuating numbers), at least one of which has a nonGaussian distribution, to generate observed "effects" or "observations". In this situation even a single neuron can infer a cause from observed effects, merely by adjusting its synaptic weights to parallel a column of the mixing matrix (row of the inverse), using Hebbian learning. This model has been described as 'the simplest, cleanest, and thus most robust learning scenario imaginable' ((Elliott, 2012)).
.
Provided that at least one of the sources has a nonGaussian distribution, these input vectors can exhibit both second-order and higher-order correlations. In our model each agent is a single "neuron" which receives the same input vectors as every other agent, and is attempting to learn weights that permit it to track the only source which is nonGaussian, by responding to the higher-order input correlations induced by mixing. If the mixing matrix is orthogonal, only higher than second order input correlations are generated, and a nonlinear Hebbian rule can always learn the appropriate unmixing weights (Hyvärinen and Oja, 1998). In this situation cooperative learning, via agent interaction, is not needed, since individual agents can always successfully learn. However biologically even this elementary task cannot always be achieved, because (1) residual misleading second-order statistics may be present (mixing is non-orthogonal, e.g. because of inadequate preprocessing) and (2) the learning rule might not be completely synapse-specific. We and others have shown that under these conditions spurious or irrelevant solutions can be learned, especially if the starting weights are unfavorable. If weights are initialized randomly, only those agents that happen to start close to the correct solution may succeed (Cox and Adams, 2014; Elliott, 2012). However, if agents could learn from each other, in a supervised manner, as well as directly from the input data, then successful learning by one agent might seed learning by other agents. One could refer to these 2 learning styles as "discovery" (private individual learning directly from observations) and "education" (public learning by observation of discoveries). A key to human cognition might be the harnessing of individual discovery to collective learning (Boyd et al., 2011) (Adams and Cox, 2012). While local minima/plateau phenomena are particularly transparent in the ICA model ((Rattray, 2002)) (Elliott, 2012), they can hinder most other more sophisticated unsupervised learning models. Our results suggest that communication is unlikely to be a panacea for suboptimal learning: populations can get trapped in plausible but parasitic myths. We briefly discuss possible ways to ameliorate this Achilles heel of collective learning.

# METHODS

**Unsupervised learning with ICA**

The problem in ICA is to unmix a set of independent unit-variance sources that have been mixed by a mixing matrix

**x** = **Ms**

where **x** is the observable vector of mixed inputs, **s** is a vector of hidden independent sources, and **M** is the mixing matrix. In all our simulations M was orthogonal and there were only two sources, one with a Gaussian distribution and the other Laplacian. This ensures that, in the absence of crosstalk, there is only one learnable solution, a synaptic weight vector pointing in the same direction as the column of the mixing matrix associated with the nonGaussian source (Rattray, 2002). For simplicity we only consider one output neuron which tries to find a weight vector that will 'unmix' the sources so that the output neuron tracks one of them (see figure 1a).

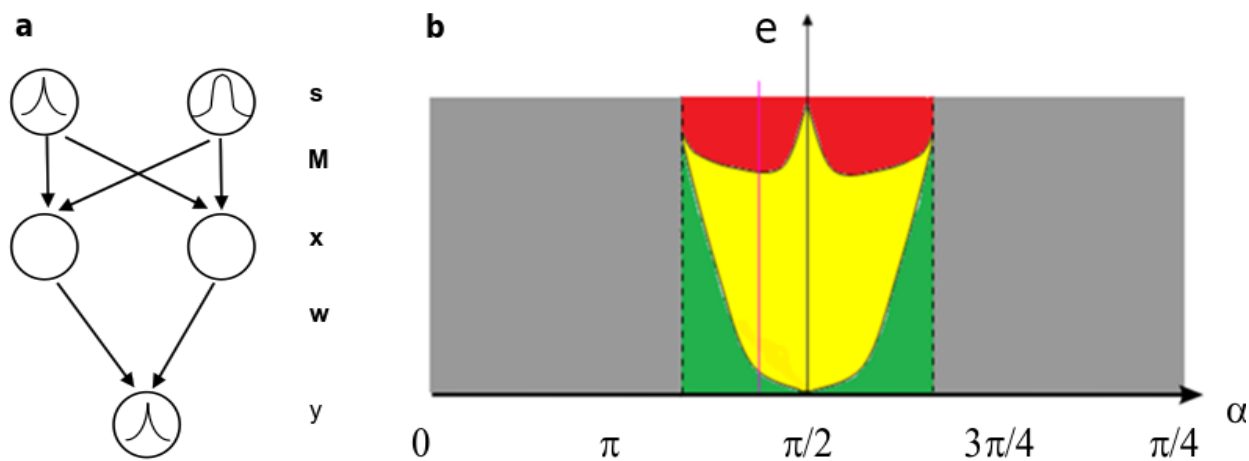

Figure 1   Panel **a**   A source vector **s** is mixed by an orthogonal mixing matrix **M** generating an input vector **x**. An unmixing vector **w** is sought that will produce an output y that will track the left hand, nonGaussian source.
Panel **b**   Single agent basins of attraction for the PC and the IC. This shows the fixed points of the learning rule as a function of the crosstalk parameter e (ordinate) and the angle $\alpha$ For e/$\alpha$ pairs that lie in the green area, there is only one stable fixed point, the IC, and in the red area only the PC. In the yellow area there are two fixed points, the IC and the PC. In the grey area there is a smooth change (a mixture) from the IC to the PC as e increases. The dotted vertical line marks the critical angle $\alpha$ below which there is no possibility of a crosstalk threshold, which in our simulations is 69° (for the cubic nonlinearity). The pink line represents an example of a randomly chosen example of **M** where $\alpha$ is between 69° and 90°. As the dimension of the network gets $\alpha$ will likely become closer and closer to 90° (see text).

We use the one unit negentropy-maximizing rule (Hyvärinen and Oja, 1998)

$\Delta$**w** =   k σ **x** f(**w**$^T$ **x**), normalize **w**

where k is a learning rate, **w** is the weight vector, y is the output, and σ is either +1 or -1 depending on the nonlinearity and the nonGaussian source statistics (Cichocki et al., 1997). In our case, with n=2, normalization after each iteration means that the 'basin' of attraction lies on the unit circle.

Thus the output neuron sums the inputs weighted by the current weight vector w and then passes this total 'activation' through a nonlinearity f(y) (in our case either $(y)^3$ or tanh(y)). The current weight vector is then increased by an amount proportional to f(y) and finally the updated weight vector is normalized so that no individual weight can become too large. **M** is constrained to be orthogonal so that there are no pairwise correlations introduced into the input vectors **x** (i.e. the source vectors are only rotated). Even if the original mixing is nonorthogonal this condition can be achieved by suitable preprocessing. The weight vector that enables y to track the nonGaussian source is called the independent component (IC).

We only used 2 sources, however the single Gaussian source can be replaced by multiple Gaussian sources with the same outcome (Elliott, 2012).

**Errors**
Inspecificity in the update rule is introduced by modifying the update in **w** by using an error matrix **E** as described in (Cox and Adams, 2009; Cox and Adams, 2014; Radulescu et al., 2009)

$$E = \begin{bmatrix} 1-e & e \\ e & 1-e \end{bmatrix}$$

where *e* is the, typically very low, crosstalk level.

**The mixing matrix**
The mixing matrix is orthogonal and is varied by changing the angle θ that the first column of **M** makes with the vector $[1,0]^T$.

$$M = \begin{bmatrix} cos\theta & -sin\theta \\ sin\theta & cos\theta \end{bmatrix}$$

**Principal component of E**
With crosstalk, an eigenvector of **E** becomes a possible fixed point of the dynamics in addition to the IC. With non-zero *e*, the eigenvectors of **E** are the vectors [0.707, 0.707] and [0.707, -0.707] regardless of the amount of error (since **E** is a symmetric matrix). When using a tanh nonlinearity σ is -1, and with crosstalk the PC associated with the least eigenvalue is the stable fixed point of the dynamics [0.707, -0.707] and we call this the 'PC'. With the cubic nonlinearity σ is +1 and the PC associated with the largest eigenvalue is the stable fixed point i.e. [0.707, 0.707].
The angle between the first column of **M** and the PC is denoted α, and α is varied by varying θ with the maximum angle, 90°, being achieved when θ is 45°. The angle between the PC and x-axis is fixed at 135° and so the angle between the PC and the IC is (135°- θ). For example to get an angle of 90° between the PC and the IC, θ is 45°. θ is measured counterclockwise and as θ increases $\alpha$ decreases.

**Supervised learning**

We define an agent as a single neuron feed-forward neural network (as in figure 1a) that is trying to solve the ICA problem, i.e. tracking the nonGaussian source of the input data. It can do this either directly from observing input fluctuations and using unsupervised Hebbian learning or by supervised learning, i.e. by also observing the current output of another agent that may have already learned the correct solution, in addition to directly the observing input vectors.

For supervised learning, agents use the Delta rule (Widrow and Hoff, 1960). The difference between the agent's output y is compared against a teacher's output $y_T$ (in this case another agent's estimate of the current value of the nonGaussian source). In this case the output y is the dot product of the input vector and the weight vector and has not been put though a nonlinearity

$$\Delta \mathbf{w} = k \mathbf{x}(y - y_T)$$

where Δ**w** is the update in the weight and k is a learning rate. Crosstalk slows learning for the supervised agent but does not change the direction of its weight vector (Figure 3b and 3d).

To understand the network when using a cubic nonlinearity we refer to the phase diagram in figure 2b showing the possible outcomes of learning for different values of the mixing angle $\alpha$ and crosstalk parameter e, which is described in detail in (Cox and Adams, 2014; Elliott, 2012)

With crosstalk (and/or for nonorthogonal mixing; see (Cox and Adams, 2014)), in a critical range of $\alpha$, 2 fixed points exist, one corresponding to an approximate IC, the other to an approximate PC (a 'false minimum'). Depending on the values of $\alpha$ and e, these can be either stable or unstable. In the green zone, only the IC fixed point is stable, in the red zone only the PC fixed point is stable, and in the yellow zone both are stable. In the gray zone there is only one, approximate IC, fixed point which is a smooth blend of the IC and PC.

The pink line represents a particular example mixing matrix and the closer the pink line is to the y axis the further apart are the IC and the PC (in this study measured by the angle $\alpha$, with the maximum distance being 90°). The greater the dimensionality the greater the chance that a randomly chosen matrix (the pink line) is close to $\alpha = \pi/2$.

In the yellow parameter region there are basins of attraction for both the PC and the IC (see figure 3) and the relative sizes of the basins change with error (i.e. as one moves up the y axis). Qualitatively similar behaviour is seen using the tanh nonlinearity ((Cox and Adams, 2014) and Results).

To determine the basins of attraction for a particular **M** the weights were started from a particular point on the unit circle and allowed to converge (to either the PC or the IC). This was done for many points on the circle for different error rates (see Results).

**Interacting agents**

We studied situations in which 2 or 4 potentially interacting agents can switch their learning strategies between unsupervised and supervised, while all observing the same input vectors, after a variable number of iterations which defines a Duty Cycle. Each run breaks down the simulation into sections where the agents simultaneously change their learning strategy randomly. A simulation typically runs for 200,000 epochs and this will be split into for example 10 groups of 20,000 epochs each of which has a randomly assigned learning type (U, S1, S2, S3 each with probability of ¼, see below). And so in this case the 'Duty Cycle' would be 20,000 iterations. In other runs the duty was just 1.

**2 agents**
With 2 agents there are 4 possible learning setups.

Type 1 (U)
Both Agents learn on their own, i.e. unsupervised, each agent using the ICA one-unit rule algorithm to find the IC.

Type 2 (S1)
Agent 1 learns from agent 2 and agent 2 learns unsupervised. Here agent 2 does ICA but agent 1 is being taught by agent 2 using the Delta rule.

Type 3 (S2)
Agent 2 learns from agent 1 and agent 1 learns unsupervised. Here agent 1 does ICA but agent 2 is being taught by agent 1 using the Delta rule.

Type 4 (S3)
Agents 1 and 2 learn from each other using the Delta rule. In this case learning becomes ineffective.

**4 agents**
The Duty Cycle was 1 and at each iteration each agent chose randomly between supervised and unsupervised learning: the supervised learning used the Delta Rule as before but this time an average of the outputs of the other agents was used instead of the output of a randomly chosen individual agent. This "meanfield" simplification gave similar results to runs under otherwise identical conditions where an agent learned from one randomly chosen other agent.

**RESULTS**

We present our results as follows:
In Section 1 we explore the basins of attraction for this ICA model as crosstalk is varied since these data will form the PC/IC 'landscape' upon which agents evolve as they are exposed to inputs. In Section 2 we look at how two agents evolve without crosstalk (both non-interacting and interacting), and then how agents evolve on the landscape (elucidated in Section 1) with crosstalk, again both non-interacting and interacting, and compare the two situations. In Section 3 we extend the model to four agents.

**Section 1**
**1.1 One agent basins of attraction**

If independently fluctuating signals are linearly combined to form input vectors, a single nonlinear neuron-like unit equipped with Hebbian synapses can always learn weights that unmix the inputs so as to track a hidden source signal, provided the inputs are uncorrelated at second order, and the learning rule is connection-specific and correctly signed (Hyvärinen and Oja, 1998). However if the rule isn't completely synapse-specific or the inputs are not perfectly decorrelated, or both, instead of learning correct unmixing weights, the unit can learn can learn weights that do not unmix, and correspond to the principal eigenvector of EC (the product of matrices describing the pattern of crosstalk or correlations), especially if the weights start close to this incorrect solution (Elliott, 2012); see Methods. In order to test whether communication between 1-neuron agents could overcome this type of difficulty, we first characterized the behaviour of individual non-interacting agents.
 In the case of the cubic nonlinearity, an elegant mathematical analysis has delineated the fixed-point structure of learning (Elliott, 2012). As expected from this analysis, we found that depending on the initial weights either the correct ("IC") or incorrect ("PC") solution was learned. Although a tanh nonlinearity is considered to be more robust (Hyvärinen et al., 2004), no analysis of its behaviour is available, so we then explored its basins of attraction in more detail.

**1.2 Cubic nonlinearity**
An initial assessment of the IC and PC basins using the cubic nonlinearity showed that, for $\alpha = 90°$, a crosstalk level of 0.165 divided the PC and IC basins into equal halves (data not shown) so that initial weight vectors between $[-1/\sqrt{2}, 1/\sqrt{2}]$ and $[0,1]$ would converge to the IC (represented by angles 135° to 90°) and initial weight vectors between $[1/\sqrt{2}, 1/\sqrt{2}]$ and $[0,1]$ (represented by angles 45° to 90°) would converge to the PC.

**1.3 tanh nonlinearity**
Since analytical results are not available for the tanh nonlinearity, we explored the 1-agent basin of attraction in more detail. Due to symmetry, we only need to look at the fate of the initial weight vectors represented by angles between 45° (the PC) and 135° (the IC) – see the legend of Table 1. The crosstalk was varied from 0 to 0.16.

We looked at three different mixing matrices where the angle between the first column of **M** and the "PC" was 78°, 83° and 90°. The PC is [-0.707, 0.707] so for 78° the first column of **M** is [0.54, 0.84], for 83° [0.62, 0.78] and 90° [0.707, 0.707]. For these angles 2 fixed points coexisted (yellow zone in figure 1b) and the fixed point that is actually learned depends on the initial weight vector and the amount of crosstalk. In this section we characterize this dependence, by systematically varying the initial weights.

For an angle of 78° between the PC and IC as error increased above the lower threshold the PC basin increased from 0 to 30%, expanding rapidly to 100 % on reaching the upper threshold, and the IC basin shrank from 100% to 70%. For an angle of 83° the range of the PC was from 0 – 40% and the IC shrank from 100 to 60%. For an angle of 90° between the IC and PC the PC basin expanded to nearly 100% in a smooth manner before the upper threshold was reached (0.16) (see figure 3 and table 1).

|       | degrees | 15pi/20 (135°) | 14pi/20 (127°) | 13pi/20 (118°) | 12pi/20 (108°) | 11pi/20 (99°) | 10pi/20 (90°) | 9pi/20 (81°) | 8pi/20 (72°) | 7pi/20 (63°) | 6pi/20 (54°) | 5pi/20 (45°) |
|-------|---------|----------------|----------------|----------------|----------------|----------------|---------------|--------------|--------------|--------------|--------------|--------------|
| error |         |                |                |                |                |                |               |              |              |              |              |              |
| 0.0   |         | ic             | ic             | ic             | ic             | ic             | ic            | ic           | ic           | ic           | ic           | Ic           |
| 0.01  |         | pc             | ic             | ic             | ic             | ic             | ic            | ic           | ic           | ic           | ic           | Ic           |
| 0.02  |         | Pc             | Pc             | ic             | ic             | ic             | ic            | ic           | ic           | ic           | ic           | Ic           |
| 0.03  |         | Pc             | Pc             | ic             | ic             | ic             | ic            | ic           | ic           | ic           | ic           | Ic           |
| 0.04  |         | Pc             | Pc             | Pc             | ic             | ic             | ic            | ic           | ic           | ic           | ic           | Ic           |
| 0.05  |         | Pc             | Pc             | Pc             | ic             | ic             | ic            | ic           | ic           | ic           | ic           | Ic           |
| 0.06  |         | Pc             | Pc             | Pc             | Pc             | ic             | ic            | ic           | ic           | ic           | ic           | Ic           |
| 0.07  |         | Pc             | pc             | Pc             | Pc             | ic             | ic            | ic           | ic           | ic           | ic           | Ic           |
| 0.08  |         | Pc             | Pc             | Pc             | Pc             | Pc             | ic            | ic           | ic           | ic           | ic           | Ic           |
| 0.09  |         | Pc             | Pc             | pc             | pc             | Pc             | ic            | ic           | ic           | ic           | ic           | Ic           |
| 0.10  |         | Pc             | pc             | pc             | Pc             | Pc             | ic            | ic           | ic           | ic           | ic           | Ic           |
| 0.11  |         | pc             | pc             | pc             | pc             | pc             | pc            | Ic           | Ic           | Ic           | Ic           | Ic           |
| 0.12  |         | pc             | pc             | pc             | pc             | pc             | pc            | Ic           | Ic           | Ic           | Ic           | Ic           |
| 0.13  |         | pc             | pc             | pc             | pc             | pc             | pc            | pc           | pc           | Ic           | Ic           | Ic           |
| 0.14  |         | pc             | pc             | pc             | pc             | pc             | pc            | pc           | pc           | pc           | pc           | Ic           |
| 0.15  |         | pc             | pc             | pc             | pc             | pc             | pc            | pc           | pc           | pc           | pc           | Ic           |
| 0.16  |         | pc             | pc             | pc             | pc             | pc             | pc            | pc           | pc           | pc           | pc           | pc           |

Table 1  Basins of attraction with an angle of 90° between the PC and the IC, using tanh nonlinearity. The angles (between the starting weight vectors and the PC of E top row) span a sector of a circle from 45° to 135° (anticlockwise from the x-axis). The point at 135° is the PC and at 45° the IC. The table shows where the network converges to when initialized to a starting vector which is represented by the angle. For example when the start vector is at [-0.47, 0.88] which is an angle of 118°, and the error is 0.04, the weight vector converges to the PC (see red disc in figure 3)  but when the error is 0.02 it converge to the IC (blue disc in figure 3).

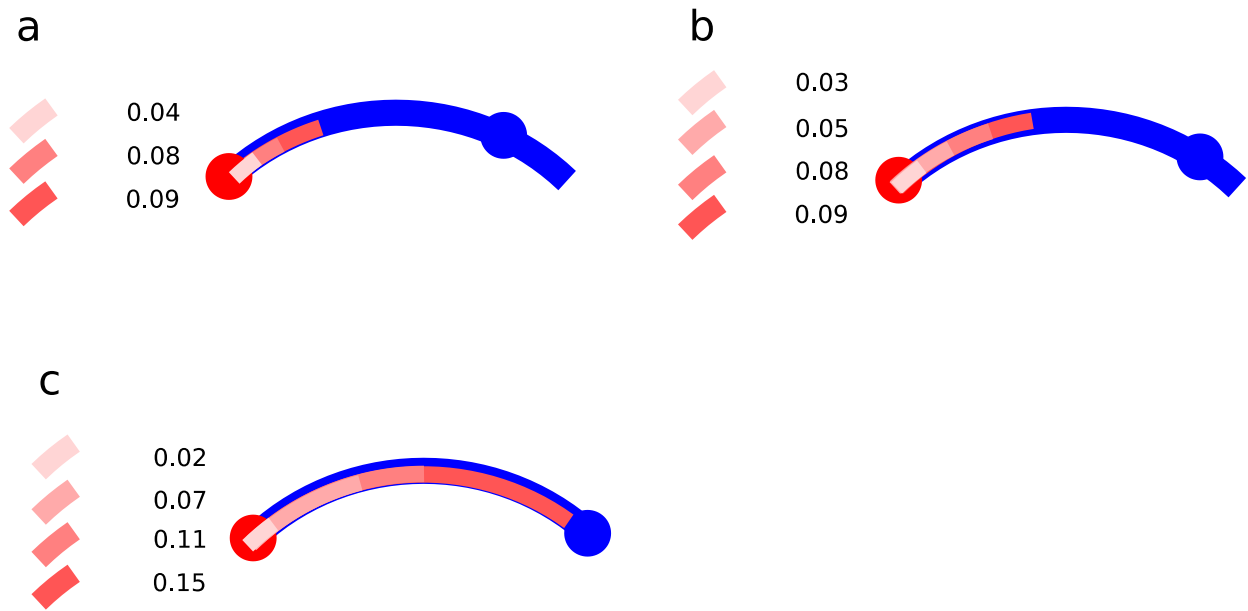

Figure 2 Basins of attraction for different initial weight vectors starting at various angles between the IC and the PC, using a tanh nonlinearity. The red disk corresponds to the direction of the PC vector, and the blue disk to 3 different directions of the IC vector relative to the fixed PC vector (alpha = 78 degrees in panel a; 84 in b; 90 in c; note that as alpha decreases from c through a, the distance between the red and blue disks decreases correspondingly). The arcs represent various weight vector starting directions. The pink arcs show the directions of starting weights that converge to the PC for various crosstalk levels, the shading corresponding to different levels. The blue arcs show the IC basin at zero crosstalk. Where the pink and blue arcs overlap, PC and IC basins co-exist. If the crosstalk level is below the indicated pink shading level, the corresponding starting weights converge to the IC, at or above that level, they converge to the PC. In a and b the upper crosstalk thresholds (above which only the PC is stable) are 0.095 and 0.09 respectively and so at the indicated crosstalk levels the PC basin did not invade the IC basin very far. In panel c, on the other hand, the PC basin expanded in a smooth way as crosstalk increased all the way up to the upper error threshold, above which the IC is never stable; see also Table 1.

Since the 90° angle case showed a smooth increase in the PC basin all the way up to the IC as error increased we decided to use this matrix to explore multi-agent learning, in the next section.

**Section 2**
**Interacting agents**

**2.1**
**'teacher' and 'student'**
An initial natural situation in when there is a 'teacher' who is at the solution and a 'student' who is not. Under these circumstances we find that the student learns ten times faster when learning from the teacher than by learning on its own, i.e. unsupervised (figure 3a and 3b). Further, if there is some crosstalk and the student is in the PC basin of attraction, then the student will never learn the IC but when learning from the teacher it will get to the IC, albeit more slowly (figure 3d).

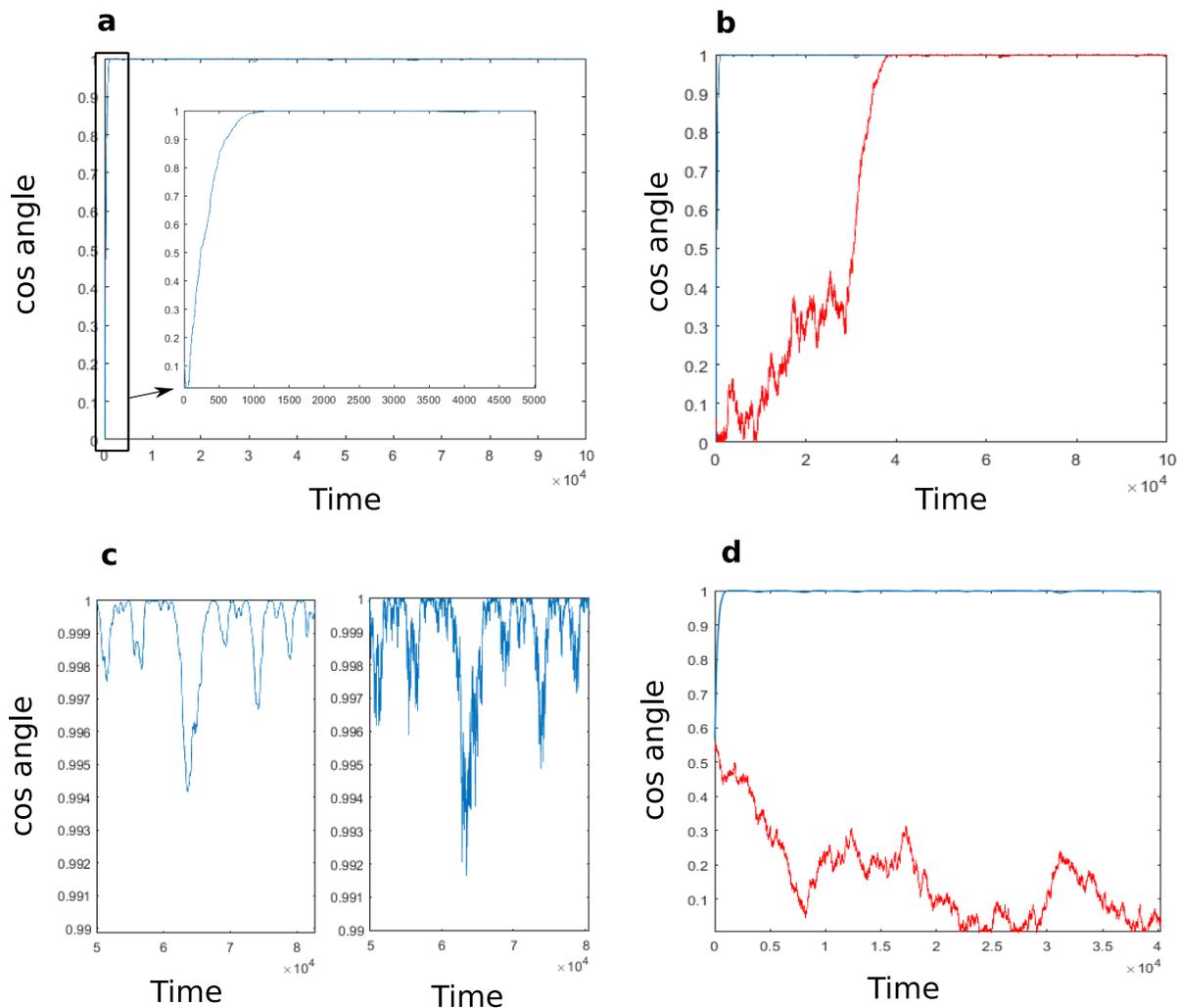

**Figure 3**   Teacher and student. Panel **a** shows the rapid progress of the student agent as it learns (zero crosstalk) using the Delta rule (with the target output being the teacher agent's output) – the inset graph is a blow-up of the first 5000 interations (rectangle). Panel **b** shows a comparison of the speed of convergence to the IC of the student agent learning (zero crosstalk) unsupervised (red trace) and supervised by the teacher (blue trace – the same one as in panel **a**). Panel **c** shows the fluctuations of the student (after reaching the IC) during both supervised (left) and unsupervised (right) learning (zero crosstalk). Panel **d** shows how the trajectories of the student agent differ when there is crosstalk and the student agent starts from the PC basin – the red trace is unsupervised and the blue trace is supervised by the teacher.

### 2.2  No crosstalk, 2 agents

The above scenario, in which where there is an agent (teacher) that only does unsupervised learning, teaching another agent (student) that only does supervised learning, appears to show that interaction between agents can help with an increase in convergence speed (figure 3b), and, even more striking, allowing the student agent to get to the IC when learning unsupervised would simply mean staying in the PC basin (figure 3d).

However, more realistically, teachers themselves must learn, and may not have found correct solutions (i.e. could equally well be at the PC), and in the following results we take an even-handed approach in which there is no privileged agent in that they all learn in the same way.

We begin with the simplest case, orthogonal mixing and no crosstalk, with only one fixed point, the IC. Several runs were done for both non-interacting and interacting agents to see if interaction had any effect on the speed of convergence for the agents. We found that the average time to convergence for the agents did not differ substantially between non-interacting and interacting agents. This was true for both cubic and tanh nonlinearities. A typical set of runs is shown in figure 4. Notice how the interacting agents synchronise their learning after about 7,000 iterations. The duty time in these runs was 20,000, but similar results were obtained for different duty times. In these runs learning types U, S1 and S2 were used (but not S3) with probability 1/3 each.

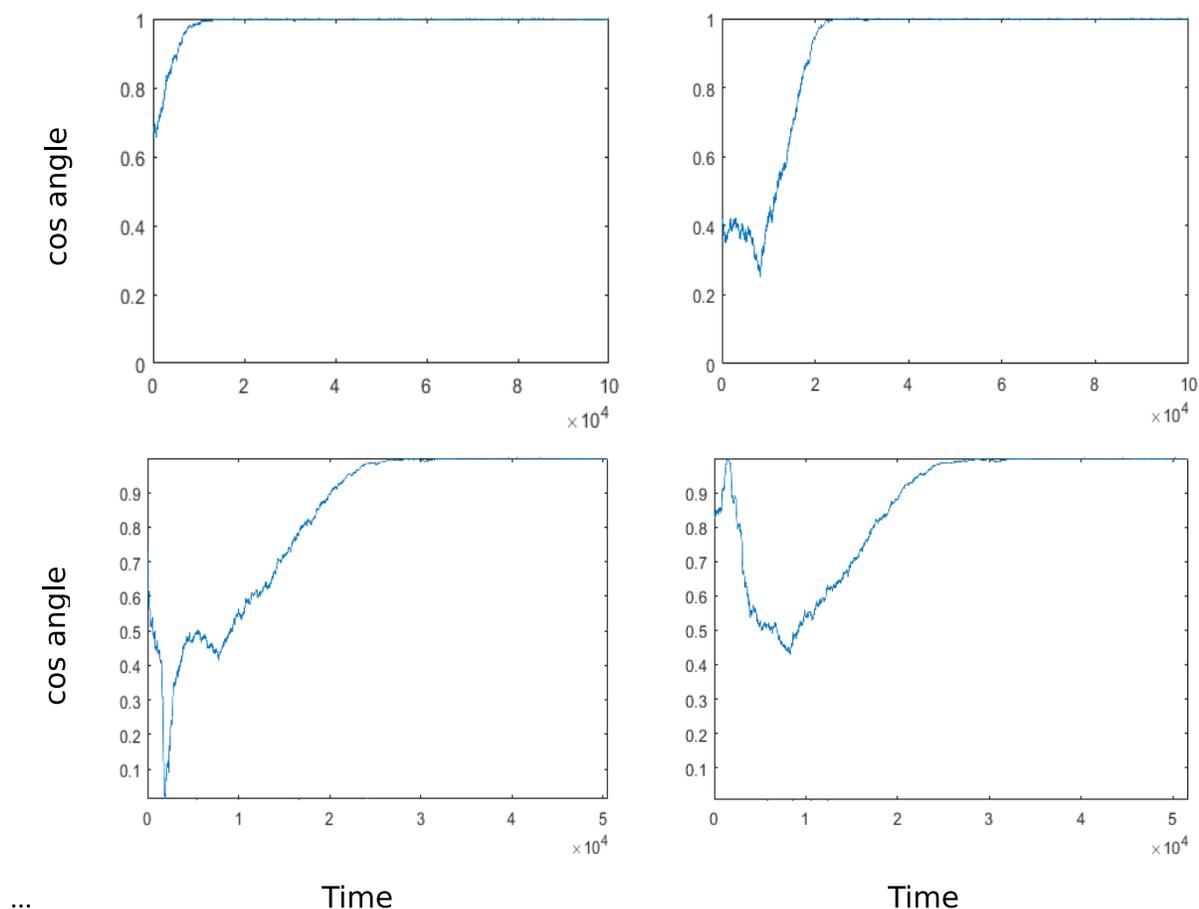

Figure 4  Non-interacting and interacting agents with zero crosstalk (tanh nonlinearity). The top 2 panels show the two non-interacting agents; the cosine of the angle the weight vectors of agent 1 and agent 2 make with the IC are shown. The weight vectors start from random values and thus converge to the IC at different times. The bottom 2 panels are interacting agents which again both also start from random weight values. Note that the interacting agents synchonise after around 10,000 interations.

## 2.2 Two Interacting agents with crosstalk

With crosstalk (but still orthogonal mixing) there are now 2 fixed points for each agent to converge to, the PC and the IC. When each agent is learning unsupervised then it will converge to the fixed point of the basin in which it initially started. However, when the agents are interacting there are further possibilities: both could end up at the IC, both at the PC, or one at the PC and one at the IC. Factors that are likely to affect the outcome are the length of the Duty Time and the initial starting points of the weight vectors. The following results explore these possibilities.

### 2.2.1 Cubic

Initially we studied interacting agents using a crosstalk level of 0.165 and initial weight vectors that put agent 1 in the IC basin and agent 2 in the PC (i.e. either side of 90°). A typical run is shown in figure 5. In figure 5 the 200,000 epoch run was broken up into ten bouts of 20,000 iterationss according to the learning types U S1 U U S3 S2 S2 S2 S1 U. In figure 5 agent 1 eventually pulls agent 2 into the IC basin after three consecutive phases of S2 after initially being pulled towards the PC by agent 2 during the earlier S1 phase. When both agents are simultaneously learning from each other (S3 starting at 80,000 epochs) then the components of the weight vector of agent 1 will move towards the weights of agent 2 (and vice versa) until they are the same.

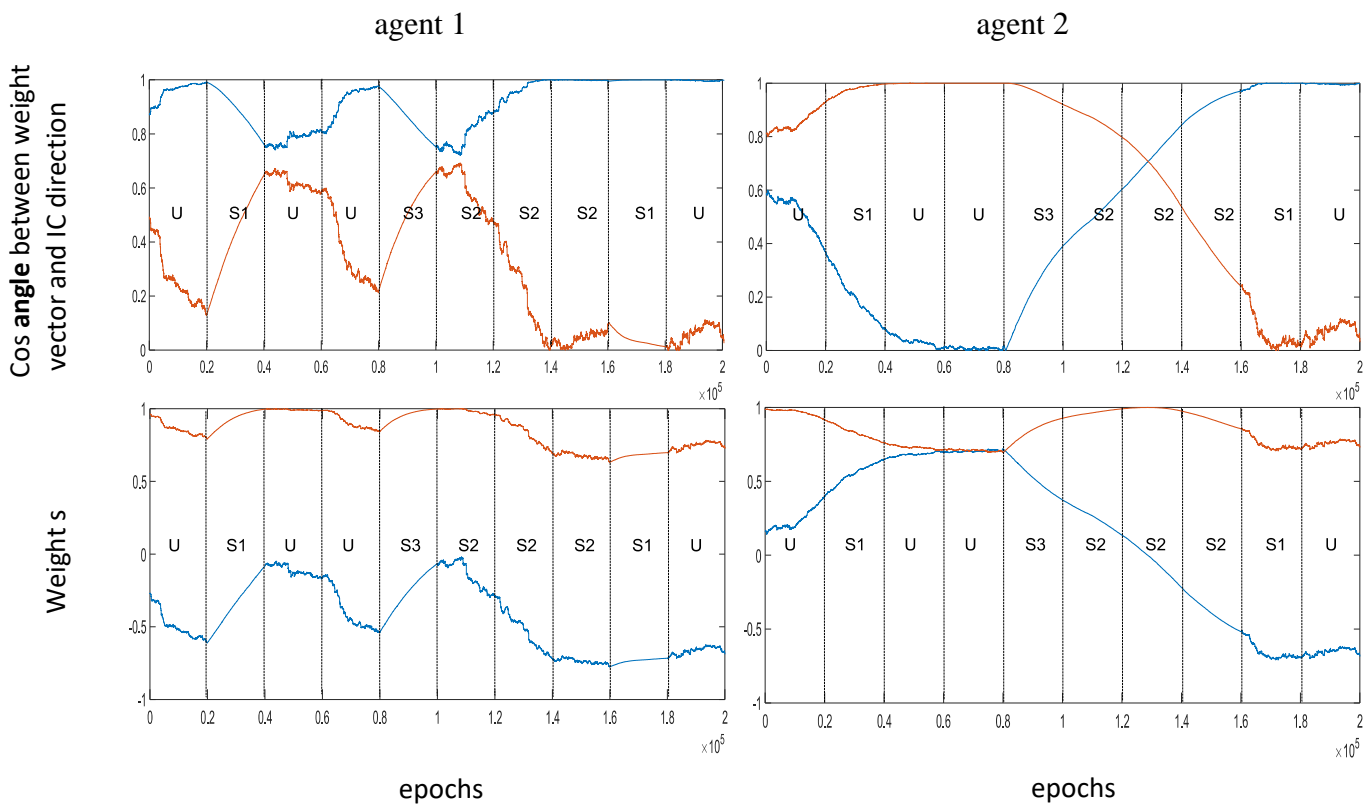

Figure 5  2 interacting; cubic nonlinearity. The top left and right plots are of the cosine of agent 1 and agent 2 weight vectors with the IC (blue) and PC (red). The lower two plots are of the weight vectors for agent 1 (left) and agent 2. Learning types switch after every 20,000 epochs according to U S1 U U S3 S2 S2 S2 S1 U.

We then examined agent interaction using the tanh nonlinearity since it is a more robust learning rule (Hyvärinen et al., 2004) and also because we wanted to explore the nature of the basins using a different nonlinearity. However no mathematical analysis of the tanh case is available.

**2.3 Tanh nonlinearity**
For all the experiments in this section an angle of 90° was chosen between the PC and the IC. For figures 6 and 7 runs were 200,000 epochs split into sections of 20,000 iterations with random reassignment between the 4 possible learning scenarios. The error rate was fixed at 0.05.

In the example below (figure 6) the initial weight vectors for both the agents were close to [0,1] which is on the 50:50 basin threshold. Both agents finally converged on the PC in this run.

# Example of both agents converging on the PC

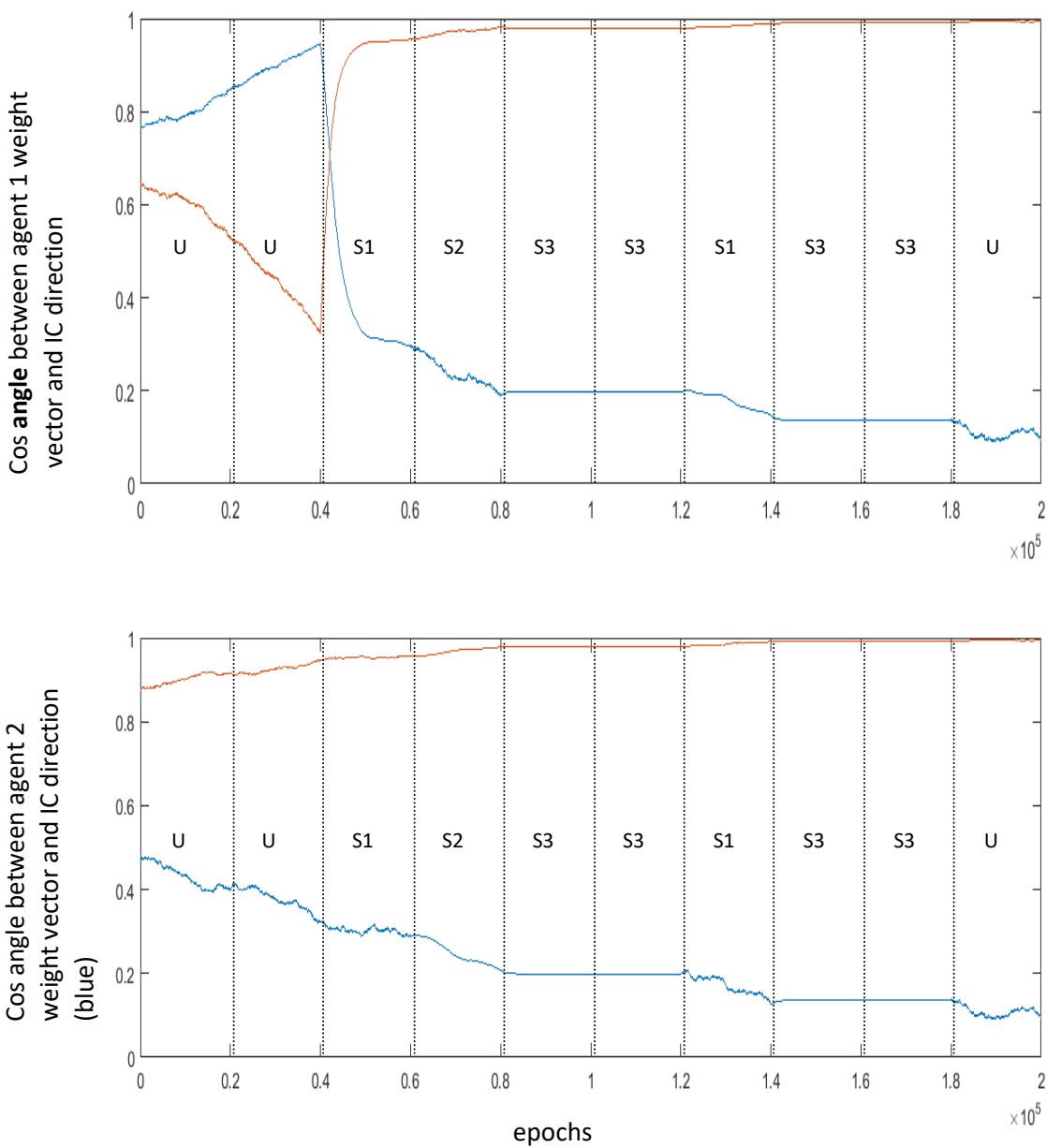

Figure 6
200,000 epoch run with agent 1 and agent 2 interacting. The vertical lines show the periods (each of 20,000 epochs) that the agents are in each learning mode. Agent 1 is the top graph and agent 2 the bottom. The error rate was 0.05 and agent 1 started in the IC basin and agent 2 started in the PC basin. The blue plots are the angles of the weight vector with the IC and the red plots the angle with the PC. Here the agent in the PC basin drags the agent starting in the IC basin into the PC solution.

In figure 6 the 200,000 epoch run was broken up into ten bouts of 20,000 epochs according to the learning types U U S1 S2 S3 S3 S1 S3 S3 U.

The initial weights were [0.0878, 0.9961] (about 85°) for agent 1 and [-0.2825, 0.9593] (about 106°) for agent 2 which, for the error rate of 0.05, puts agent 1 in the IC basin and agent 2 in the PC basin (see Table 1). Two bouts of unsupervised learning allowed each agent to progress towards their respective solutions (agent 1 the IC and agent 2 the PC). Then at 40,000 epochs each agent switched to S1 learning where agent 1 learns from agent 2 (which has already made good progress in learning the PC) resulting in agent 1 rapidly changing to learning the PC. After 20,000 epochs of S1 learning both agents were in the PC basin and the type of learning that followed did not affect the eventual fate of each agent, namely converging on the PC solution. As expected when both agents are learning from each other (i.e. S3), the components of the weight vectors move towards each other - if the components of the weight vectors are agent 1 and agent 2 are already (due to previous learning) close to each other then very little change is seen.

In the next example (figure 7) the 200,000 epoch run was broken up into ten bouts of 20,000 epochs according to the learning types U S2 S2 S1 S2 S1 S3 S2 U U.

## Example of both Agents converging on the IC

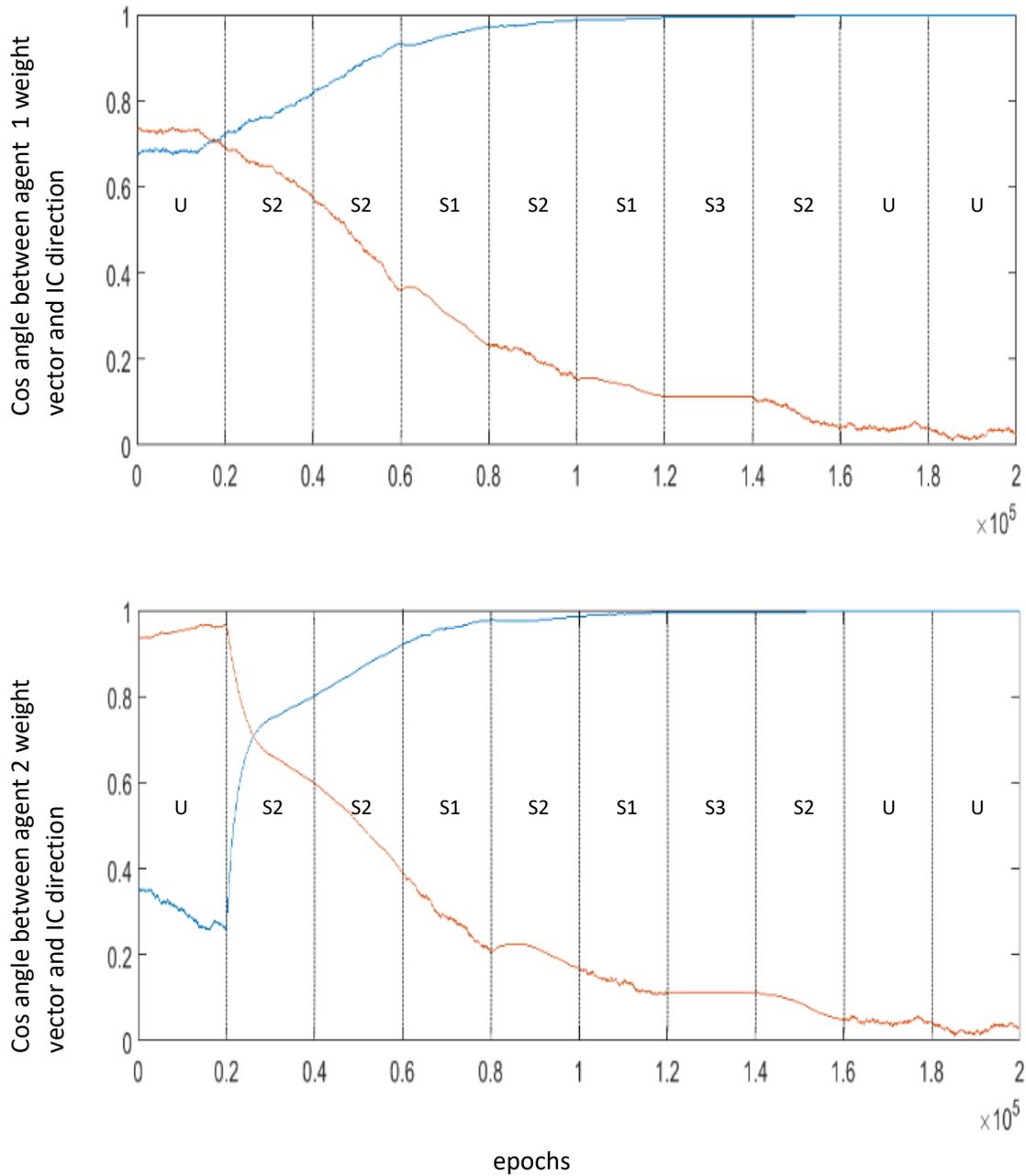

epochs

Figure 7
200,000 epoch run with agent1 and agent 2 interacting. The vertical lines show the periods (each of 20,000 epochs) that the agents are in each learning mode. Agent 1 is the top graph and agent 2 the bottom. The error rate was 0.05 and agent 1 started in the IC basin and Agent 2 started in the PC basin. The blue plots are the angles of the weight vector with the IC

The initial starting vectors were agent 1 [-0.0382   0.9993] and agent 2 [0.4078   0.9131]  which from Table 1 shows that agent 1 started in the IC basin (about 92°) and agent 2 started in the PC (about 113 °) basin.

In the run shown in figure 7 agent 1 begins in the IC basin and agent 2 is in the PC basin. For the first 20,000 epochs both agents learn unsupervised resulting in agent 1 moving closer towards the IC and agent 2 moving closer towards the PC. The second batch of 20,000 epochs has each agent learning via S2 learning which results in agent 2 being dragged out of the PC basin and into the IC basin. As soon as both Agents are in the IC basin they stay there and eventually both converge to the IC solution.

For a certain crosstalk level what an agent will learn depends on the initial weights of the network, i.e. whether it is in the basin of attraction of the PC or the IC. If one agent starts in the PC basin and the other in the IC basin then depending on the particular sequence of learning cycles (supervised or unsupervised) then each agent can pull the other agent into its basin of attraction. We looked at the case where the IC was as far away as possible from the PC (90°) which we show allows a crosstalk level (0.15) that splits the basins 50:50 so that there is an equal chance for an agent to learn the IC or PC (since it has an equal chance of starting in the PC or IC basin). If the agents did not interact then each agent would half of the time find the IC and half the time the PC.

**Agents starting from PC and IC, duty time = 1**

We then studied interacting agents from the extreme viewpoint of starting the interacting agents at the PC or IC (one exactly at the PC and the other exactly at the IC) and then studying their trajectories at various crosstalk levels. The duty cycle was set at 1, i.e. the extreme condition where the agent changes learning type randomly at each iteration. Data was collected over many hundreds of runs for two different learning rates, 0.0002 and 0.0001 – see figure 8.

For every run, both agents always ended up in either the PC basin or the IC basin, depending on the crosstalk level. As expected at low crosstalk, both converge to the IC, and at high crosstalk both converge to the PC, while for intermediate crosstalk, both can converge to either the IC or the PC, though this intermediate range narrows as the learning rate decreases. Presumably with extremely slow learning, there is an abrupt transition from both-IC to both-PC.

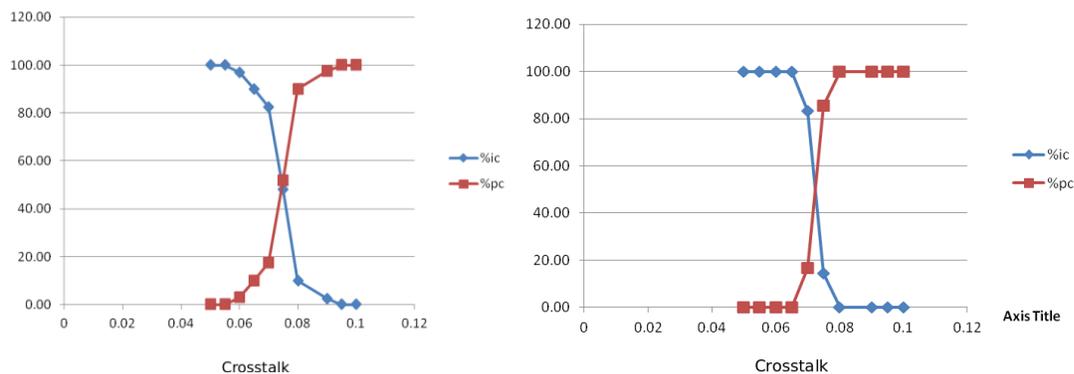

Figure 8    Fate of interacting agents as crosstalk is varied. One agent was started at the IC and the other agent at the PC. Each run was for 200,000 iterations and at each iteration both agents changed learning type. The left hand graph has a learning rate of 0.0001 and the right hand graph 0.0002. The plots show the percent of time both agents went to the IC or PC at different crosstalk levels.

As the learning rate was reduced the fate of the agents became more binary in that for a particular crosstalk level they would both go to either the PC or the IC. For instance take crosstalk at 0.065: for a learning rate of 0.0002 10% of the time both agents would go to the PC but at a learning rate of 0.0001 they both always went to the IC.

**Section 3**

### 3.1   4 interacting agents, zero crosstalk

With 4 agents a modified program was used in which when an agent was doing supervised learning, it would use the average of the output of the other agents in the delta rule (since otherwise the number of learning types becomes very large). Crosstalk was zero (so the only fixed point was the IC) and agents starting with random weight vectors and the average time it took for the agents to converge compared. The Duty Time of 1 as in the previous section.. From our limited number (40 unsupervised and 100 supervised) of runs we found that the average time to convergence was 1.5 times slower for interacting agents. As with the two agent case, all four agents at some point synchronized and travelled as a 'group' to the IC.

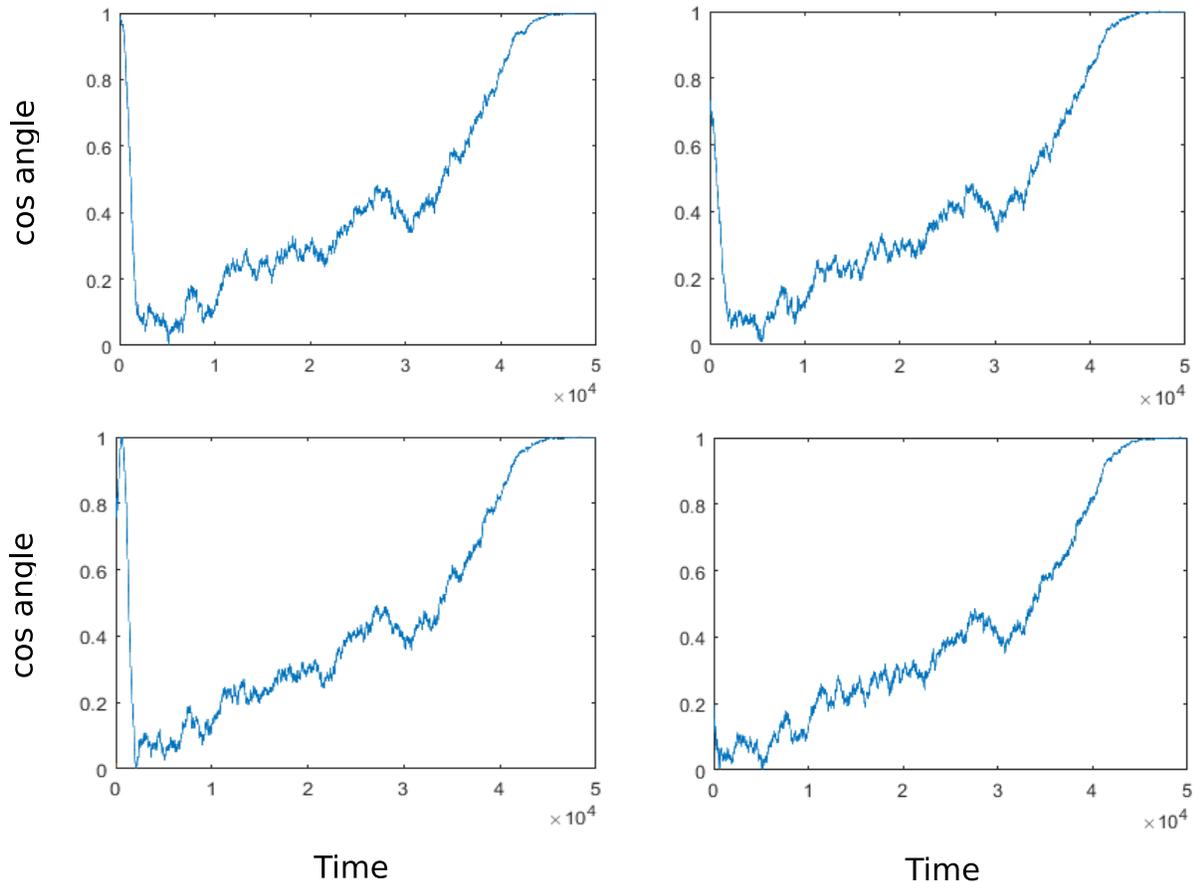

Figure 9   A typical example of 4 interacting agents without crosstalk. The above plots are of each agent as it converges to the IC with the y-axis the cosine of the angle of the weight vector with the IC.  Each agent started from random weights and after each iteration changed randomly to either unsupervised learning or supervised learning. The number of iterations was 50,000. Note how after only about 2,000 iterations all 4 agents began to synchronise their learning.

## 3.2  4 interacting agents with crosstalk

Increasing crosstalk from zero changes the sizes of the PC and IC basins so we looked at the case with crosstalk = 0.05 to see if different basin sizes would alter the convergence times of U and S. Under these circumstances in the non-interacting learning mode the agents will simply converge to bottom of the basin they start in. In the interacting learning mode the situation is more complicated since where all the agents end up depends on the position of each agent's weight vector in the basin it starts in as well as the number of agents in each basin at the start of the run. We found that, as in the 2 agent case, when interacting all 4 agents either went to the IC or PC, with no mixed outcomes. Synchronisation of the agents also occurred (at around 1,000 iterations in Figure 10).

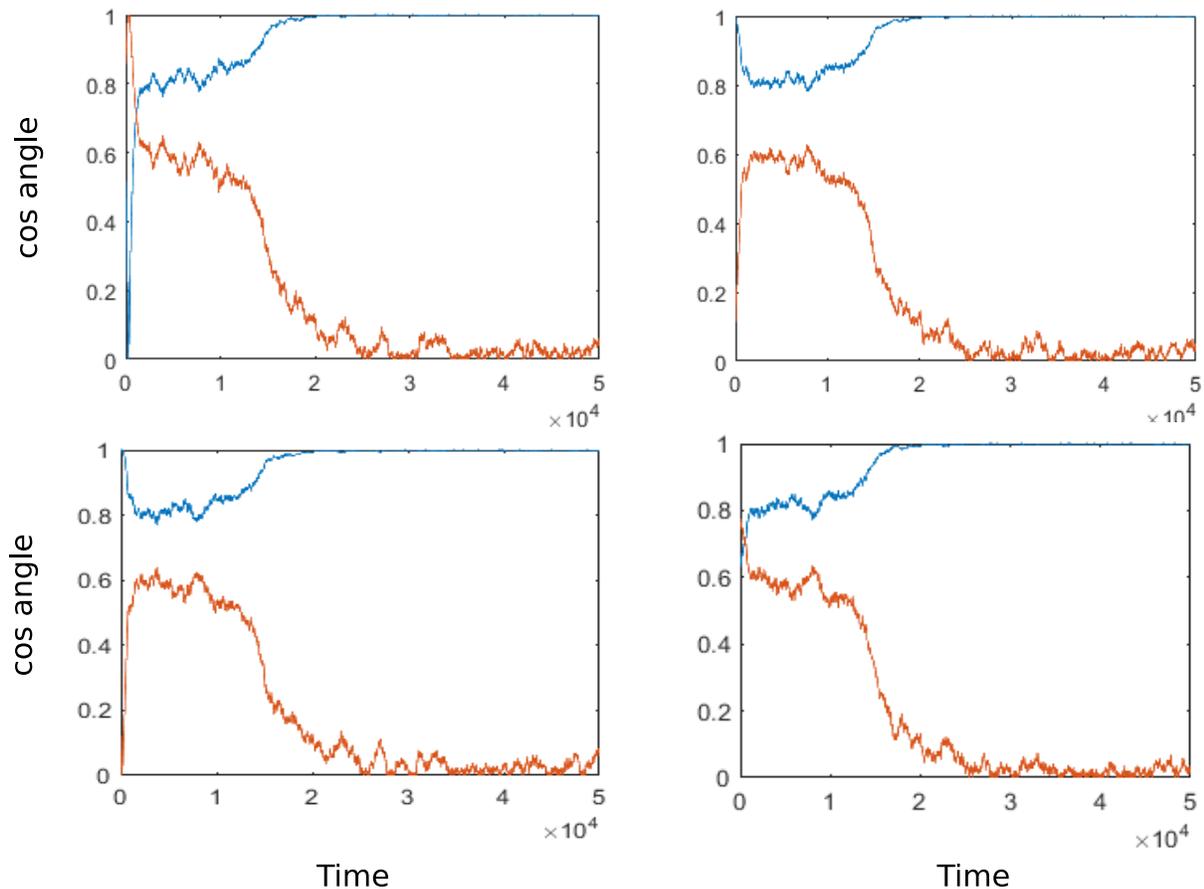

Figure 10
A typical example of 4 agents interacting with crosstalk (0.05). In this case all the agents went to the IC (blue line) and began moving as a group at around 5,000 iterations.

**DISCUSSION**

How animals, especially those living in social groups, learn about the world is of relevance to understanding how Homo *sapiens* and culture have evolved. Survival strategies can be coded into the genome or learned through experience, and learning can be done independently or by instruction. The machinery, such as symbolic communication, selective attention, imitation, cooperation, trust etc is of great interest, but we bypass these issues by assuming perfect attention and communication.

Information is probably stored in brains by synaptic strength changes (Bartol et al., 2015), but it seems unlikely that social learning is accomplished by direct copying of synaptic weights, not least because an individual does not have access to her own weights. It seems more likely that social learning employs essentially the same mechanisms as individual learning, in particular close observation of the physical environment, but with an additional, supervisory, component (Bengio, 2012). In our model, using the delta rule, this is explicit: supervised learning is driven by output, input and an explicit (though possibly incorrect) target. Learning is often slow, difficult or unsuccessful, or incorrect solutions may be learned, even when unlimited amounts of relevant data are available.

In the ICA model a 'hard' problem can be thought of one in which the IC basin of attraction is small and so without crosstalk (or second order correlations) all problems are 'easy' since only the IC basin exists. With crosstalk and/or color a competing PC basin appears, and if this basin increases in size (the IC basin decreases in size), the probability of starting near the IC may be lowered. Deep learning can be slowed by the existence of plateaus and/or false minima in weight space (Atakulreka and Sutivong, 2007; Bengio, 2009; Erhan et al., 2010; LeCun et al., 2015; Saxe et al., 2013; Wessels and Barnard, 1992) and in this related sense the problem is hard. One important objective of deep learning is to get into a favourable part of weight space where gradient descent can be effective. For our model utilizing a lucky agent that is in a 'good' part of weight space (IC basin) to teach other agents in 'bad' parts (PC basin) makes the problem easy for the latter agents.

In our model both individual and social learning are driven by observation of the mixed input vectors, but individual learning is unsupervised, with no access to the underlying source values, while social learning is also driven by access to, possibly incorrect, estimates of underlying source values. In a population of interacting agents, some individual learners will learn veridical solutions (e.g. IC) while others will learn misleading (e.g. PC) solutions. The larger the population the more likely it becomes that at least one agent will unsupervisedly learn the correct solution, and one might hope that this individual can then seed correct learning by the whole group. However, our results suggest social learning is a double-edged sword, equally facilitating veridical and misleading learning. A related situation underlies the "Rogers Paradox" (Rogers, 1988) (Boyd and Richerson, 1995) (Boyd et al., 2011). In Rogers's evolutionary game model, reproducing agents can learn 2 alternative behaviors, only 1 of which confers a fitness advantage in the current environment. They can learn these behaviors either individually or socially. As a result, the degree of fitness advantage conferred by social learning depends on the proportion of the population that uses individual learning - when most of the population is composed of individual learners (who always match their behavior to the current environment), social learning provides an advantage, but as the proportion of social learners increases, learning mismatched behavior becomes more likely. Rogers shows that the population will evolve to contain a mix of individual and social learners such that average fitness is identical to that of pure individual learners, so that social learning confers no advantage. Similarly, in our model social learning does not selectively accelerate veridical learning. In both cases one must postulate that veridical learning confers a further advantage - in our case the ability to track a source might increase supervisory effectiveness. The IC allows inference which is useful in guiding behaviour whereas the PC describes the data compactly but does not reveal underlying causes and thus gives a lower fitness advantage (Field, 1989). Of course ICA is a linear mixing model and the real world is not, but doing ICA on e.g. natural images could make progress towards finding the causes (Bell and Sejnowski, 1997; Chen and Gopinath, 2000).

Indeed it has been suggested (Nicholls et al., 2012) that the goal of sensory neocortical circuitry is to infer causes: to convert sensations to perceptions. One possible iterative approach is using successive layers of ICA-like learning followed by alternating layers of marginal Gaussianization and whitening (Chen and Gopinath, 2000; Laparra et al., 2011; Lyu and Simoncelli, 2009). Clearly such a deep, iterative strategy would be imperilled by failure of the ICA-like step, for example because of synaptic crosstalk or imperfect whitening, and such failure would be more likely for hard problems. More generally it's likely that the learning that allows improved inference requires connection-specific learning. Our simplified model may throw light on these issues, especially in the context of social learning.

We have studied a minimal model of learning in which two agents are able to learn a problem in either unsupervised or supervised fashions. In this model each agent learns a task using a neural network involving incremental weight changes based on data from the environment (unsupervised ICA) or also using information from another agent (supervised using the delta rule). We look at the simplest possible case where the observed data are generated from a mixture of two independent sources one of which is Laplacian and the other Gaussian. In the problem we study, multiple interacting agent ICA learning, the data are usually assumed to be 'white' (meaning the data have no pairwise correlations) and perfect ICA learning will always find the IC (Hyvärinen et al., 2004) so that the behaviour of the underlying nonGaussian "cause" can be accurately inferred. In real life the ability to correctly infer the cause of observations (e.g. a tiger in the grass) will confer fitness advantages. In the real world individuals cannot reliably learn for a variety of reasons, including (a) imperfect whitening and (b) crosstalk (i.e. an imperfect learning rule).

It has been shown previously that the simple picture of a neural network learning the IC becomes more complicated when the learning rule used (the Hebb rule) is no longer exactly specific and/or the whitening is imperfect (Cox and Adams, 2009; Cox and Adams, 2014; Elliott, 2012) ; if the weight changes leak slightly to other connections (synapses) then it is possible for the network to learn the eigenvectors of **EC** where **C** is the correlation matrix (in the white case **C** = **I**, as in this paper). In the current work we use perfect whitening but allow crosstalk. However, it's likely that similar results would be found if imperfect whitening, either rather than or in addition to crosstalk, was a problem: learning can get stuck at the wrong solution if the starting weights are unfavorable (see equation 8 in (Cox and Adams, 2014; Elliott, 2012)). Note that we are defining the correct solution as one that allows veridical tracking of the underlying source fluctuations. The underlying assumption is that identifying true causes is useful, and confers a fitness advantage. Many neural network theorists instead use a different criterion: learning is deemed successful if it allows the brain to spontaneously generate (via an internal generative model) outputs that have the same statistical regularities as the real word (Hinton, 2007). In this narrow sense PC learning might be useful, because (especially when input statistics are Gaussian), the output of a few neurons that can represent the first few principal components of the input data can quite successful reconstruct the entire input. However, these outputs would not track the source fluctuations.

Our preliminary studies of 4 interacting agents suggests that one can probably extrapolate our results to the case of large numbers of agents, in situations where a slight majority of agents' initial weights lie in the IC basin of attraction. A particular agent that engages in supervised learning would then be slightly more likely to move to the IC, and as more agents find the IC, the overall movement to the IC would accelerate. This would be analogous to the meanfield ferromagnet model, where if even a tiny majority of spins align, there is a cascading movement to the majority direction. Of course once all the agents reach either the IC or PC they will remain there, since when an agent shifts to supervised learning, it will be kept at the fixed point.

We have previously noted that the role of crosstalk (i.e. the limited synapse specificity of connection strength updates) is analogous to that of polynucleotide copying errors in Darwinian evolution. In both cases the ability to acquire new information about the environment is limited by the accuracy of the elementary learning process (base replication or synapse strengthening) (Eigen, 1971). In the case of Darwinian evolution, this limit imposed on evolution by asexually-reproducing prokaryotic organisms is greatly relaxed in sexually-reproducing eukaryotic organisms, which store much more information in their genomes (Ridley, 2000), though the exact mechanisms by which sex supercharges evolution remains controversial (Otto, 2009). In particular sexual reproduction, which is essentially a protocol for the exchange of genetic information between members of the same breeding population (a "species"), can easily be exploited by "selfish genes". A well-known example is mitochondria. Selfish male mitochondrial genes that act to destroy female mitochondria (and vice versa) would undermine organismal fitness while promoting their own proliferation. Nuclear genes avoid this conflict by eliminating all male mitochondrial genes by both pre- and post-fertilization processes (Ridley, 2000) (Hurst, 1995). More generally sexual reproduction affords many opportunities for parasitic genes to flourish at the expense of organismal fitness, and sexually reproducing organisms have developed multiple mechanisms preventing free-riding. (Butcher and Deng, 1994; Hurst, 1993).

Our results hint that an analogous situation exists for social learning. While social learning, by supervision by others, can accelerate the spread of individual discoveries (see figure 3), it can equally assist in the spread of misleading ideas (in our model, the PC), which function as parasites. It has been suggested that coupled learning by multiple agents could escape local minima in deep learning networks (Bengio, 2012), but this might suffer the same problem as we find, i.e. symmetrical communication doesn't help when teachers are also learners. In order to selectively enhance collective learning of useful weight vectors (e.g. ICs), their successful learning must confer an advantage. The learning population needs not merely a shared protocol for the exchange of information (a "language") but also shared procedures for testing the utility and veridicality of concepts developed by means of unsupervised learning from observations. An extreme example of such a procedure is "science": empirical testing and dissemination of hypotheses developed by observation and experiment. However, human societies presumably developed previous less formal, but still highly cooperative, prosocial ways of testing the value of claimed "discoveries" (Marean, 2014; Marean, 2016). In particular, observationally-based ideas that do not confer actual ability to infer underlying causes and predict outcomes, must somehow by penalized, while the adoption of observationally-based ideas that lead to successful outcomes must be enhanced. Future models of the interaction of individual and social learning should test this idea.

One possible modification of our current model that might clarify this issue would be to incorporate some sort of advantage accruing to those who successfully learn the correct solution. In the simplest case one could allow occasional 'cheating': an agent obtains a 'glimpse' of the true source values, which it could use for supervised learning. One could view this as performing empirical experiments to check the validity of one's inferences. A more interesting and subtle variation of this idea would be to allow only other agents, but not the glimpsing agent, see a measure of its success. In other words, no agent could directly even glimpse the source, but nevertheless agents would be able to display a scalar measure of their success to other agents (but perhaps not to themselves), analogous to 'reputation'. We have tried to construct a minimal model to capture the essence of how supervised and unsupervised agents interact. The model is simple but has the advantage of being mathematically transparent, and the issues that arise may be relevant more generally in more complex but less transparent models. We conclude that symbolic communication alone cannot reliably boost individual intelligence and that additional social mechanisms such as trust, reputation, cooperation and science are also necessary.

IC